\documentclass[prx,amsmath,amssymb,twocolumn,superscriptaddress]{revtex4-2}

\usepackage{graphicx, bm, physics}
\newcommand{\angstrom}{\mbox{\normalfont\AA}}
\usepackage{soul}
\usepackage{xspace}
\usepackage[dvipsnames]{xcolor}

\begin{document}
\title{Superlattice Engineering of Topology in Massive Dirac Fermions} 
\author{Nishchay Suri}
\email{nishchay.suri@nasa.gov}
\affiliation{QuAIL, NASA Ames Research Center, Moffett Field, California 94035, USA}
\affiliation{USRA Research Institute for Advanced Computer Science, Mountain View, California 94043, USA}
\affiliation{Department of Physics, Carnegie Mellon University, Pittsburgh, Pennsylvania 15213, USA}
\author{Chong Wang}
\affiliation{Department of Materials Science and Engineering, University of Washington, Seattle, Washington 98195, USA}
\author{Benjamin M. Hunt}
\affiliation{Department of Physics, Carnegie Mellon University, Pittsburgh, Pennsylvania 15213, USA}
\author{Di Xiao}
\affiliation{Department of Materials Science and Engineering, University of Washington, Seattle, Washington 98195, USA}
\affiliation{Department of Physics, University of Washington, Seattle, Washington 98195, USA}
\date{\today}

\begin{abstract}
We show that a superlattice potential can be employed to engineer topology in massive Dirac fermions in systems such as bilayer graphene, moir\'e graphene-boron nitride, and transition-metal dichalcogenide (TMD) monolayers and bilayers.
We use symmetry analysis to analyze band inversions to determine the Chern number $\mathcal C$ for the valence band as a function of tunable potential parameters for a class of $C_4$ and $C_3$ symmetric potentials. We present a novel method to engineer Chern number $\mathcal C=2$ for the valence band and show that the applied potential at minimum must have a scalar together with a non-scalar periodic part. 
We discover that certain forms of the superlattice potential, which may be difficult to realize in naturally occurring moir\'e patterns, allow for the possibility of non-trivial topological transitions.  
These forms may be achievable using an external superlattice potential that can be created using contemporary experimental techniques. Our work paves the way to realize the quantum Spin Hall effect (QSHE), quantum anomalous Hall effect (QAHE), and even exotic non-Abelian anyons in the fractional quantum Hall effect (FQHE). 
\end{abstract}

    \maketitle

\section{Introduction}

The discovery of the integer quantum Hall effect (QHE) in two-dimensional (2D) systems has led to the classification of the phases of matter by their topology~\cite{von1986quantized,liu2016quantum,cage2012quantum,klitzing1980new,laughlin1981quantized,halperin1982quantized,tong2016lectures}, which is distinct from Landau's classification based on spontaneously broken symmetries. The topological phase of matter is characterized by its Chern number $\mathcal C$ that defines the quantization of its properties such as the Hall conductance and does not change unless there is a topological transition. The topologically non-trivial phases of matter are of interest for fundamental understanding as well as for the future of technologies as they host edge modes that are robust information carriers useful in spintronic and computing applications~\cite{sato2017topological,he2019topological, vsmejkal2018topological}. Exotic topological states of matter such as the non-Abelian anyons that are neither bosons nor fermions, exhibit non-Abelian braiding statistics foundational to topological fault-tolerant quantum computation and can emerge in systems displaying the fractional quantum Hall effect (FQHE)~\cite{leijnse2012introduction,sato2017topological,he2019topological, google2023non}.

Recently there has been considerable interest in the physics emerging from the formation of a periodic moir\'e pattern created due to stacking a 2D monolayer on top of another with a small rotation or lattice mismatch. The interlayer binding potential of the moir\'e pattern acts as a natural superlattice potential forming minibands that significantly alter the properties of 2D monolayer materials leading to exciting new quantum phenomena~\cite{andrei2020graphene,kennes2021moire,balents2020superconductivity}.  In addition to moir\'e patterns, contemporary experimental techniques have provided the ability to apply an externally generated or `artificial' superlattice potential to lattices, which can be more general than the naturally occurring potential and is largely unexplored~\cite{PhysRevLett.130.196201,
PhysRevB.107.195423}.
For example, one method of creating such an artificial potential is by considering the LAO/STO (LaAlO$_3$/SrTiO$_3$) 
interface that hosts a 2D electron gas~\cite{ohtomo2004high,sulpizio2014nanoscale,pai2018physics,reyren2007superconducting,caviglia2008electric}. An atomic force microscopy (AFM) tip~\cite{cen2008nanoscale,cen2009oxide,bi2010water} or a low-energy electron beam~\cite{yang2020nanoscale} can be used to distribute the charge density and `write' arbitrary potentials which can couple to the system of interest~\cite{tang2019long,tang2020frictional,briggeman2020pascal,briggeman2020engineered,briggeman2021one}. 
Another method is by considering the twisted hexagonal boron nitride (h-BN) substrate, where an intrinsic charge redistribution generates an external superlattice potential~\cite{zhao2021universal}.
Other methods such as dielectric screening~\cite{forsythe2018band,xu2021creation} can also be used for the purpose.


In this Letter, we provide the minimal form of such an externally-applied superlattice potential to engineer topological states in massive Dirac fermions that occurs in multiple systems such as in bilayer graphene, moir\'e graphene/hexagonal boron nitride, and transition-metal dichalcogenide (TMD) monolayer and bilayer materials. For a class of $C_4$ and $C_3$ symmetric potentials, we use symmetry analysis to analyze band inversions to determine the Chern number $\mathcal C$ for the valence band as a function of tunable potential parameters. We present a novel method to engineer Chern number $\mathcal C=2$ for the valence band and show that the applied potential must have a scalar together with a non-scalar periodic part. We discover that externally applied potential allows the possibility of non-trivial topological transitions that can be difficult to achieve otherwise by only naturally occurring moir\'e pattern produced by a twist or lattice mismatch. The topologically non-trivial band can be made nearly flat by increasing the strength and the periodic length of the applied potential. Partially-filled flat bands with non-trivial Chern numbers can give rise to fractional Hall states as suggested by numerical works~\cite{PhysRevLett.48.1559,PhysRevLett.106.236804,PhysRevLett.106.236802,PhysRevB.84.155116,sheng2011fractional,PhysRevLett.107.146803,PhysRevX.1.021014,PhysRevLett.108.126805,PhysRevB.85.075128,PhysRevB.85.075116}. Moreover, there is no equivalence between the simple Landau levels and a higher Chern number such as $\mathcal C=2$ which can occur in bands. Therefore engineering a flat $\mathcal C=2$ band may lead to even more exotic and exciting FQHE physics which may not be realizable with the Haldane model or an external magnetic field having Landau Levels~\cite{peter2015topological,wang2011nearly,wang2012fractional,yao2015bilayer,chen2020tunable}. We finally discuss experimental implementation of our work which allows for the possibility to realize QSHE, QAHE and even exotic non-Abelian anyons in the FQHE. 

We introduce the general Hamiltonian describing massive Dirac fermions in a periodic potential in Sec.~\ref{sec:MassiveDirac}. We consider the $C_4$ symmetric potentials in Sec.~\ref{sec:C4} and show that both the first and second star of Fourier components of the applied scalar potential $V_0(\bm r)$ are needed to induce $\mathcal C=\pm 1$ for the valence band. Subsequently, we show using symmetry analysis that $\mathcal C= \pm 2$ can only be realized in the valence band by considering both the  scalar $V_0(\bm r)$ together with a non-scalar $V_1(\bm r) \sigma_z$. We then consider the $C_3$ symmetric potential in Sec.~\ref{sec:C3} and similarly show that the scalar together with a non-scalar potential can induce $\mathcal C=\pm 2$ for the valence bands. Finally, we discuss the experimental realizations of the theory.

\section{Massive Dirac fermions in a periodic potential}
\label{sec:MassiveDirac}

The low energy physics and topological transitions in 2D materials can be well understood by considering the massive Dirac structure of the band dispersion at the high symmetry points. We consider a massive Dirac fermion under the influence of an external superlattice potential. We expand the tight-binding Hamiltonian at each valley $\pm \bm \kappa$ to obtain
\begin{align}
H_{\tau,\bm k} &= H^D_{\tau,\bm k} + V(\bm r), & H^D_{\tau,\bm k} = at(\tau k_x \sigma_x + k_y\sigma_y) + \frac{\Delta}{2}\sigma_z,
\end{align}
where $a$ is the atomic lattice constant, $t$ is the hopping parameter, $\tau$ is the valley index, $H^D_{\tau,\bm k}$ is the Dirac Hamiltonian, and $V(\bm r)$ is the external superlattice potential. For example, in the case of TMDs, the Pauli matrices describe the $d$-orbital basis $\{\ket{d_{z^2}},1/\sqrt 2(\ket{d_{x^2-y^2}}+i\tau\ket{d_{xy}})\}$.

Since the entire system is time reversal symmetric, the valley Chern number $\mathcal C_{\bm \kappa}=-\mathcal C_{-\bm \kappa}$. The strong spin-orbit coupling in TMDs leads to valley-polarized spins. The two degenerate valleys will be equally populated and a non-trivial valley Chern number will induce the QSHE~\cite{PhysRevLett.108.196802, PhysRevB.107.075424}. However, if the bands at both valleys are nearly flat and are half filled, the Stoner instability will lift the spin degeneracy, separating spin up from the spin down states. Consequently, only one of the valleys will be populated, which can possibly induce QAHE as well as FQHE if the bands are made sufficiently flat. We shall only consider the valley $\tau = +1$, referring to the valley Hamiltonian $H_{\tau,\bm k}$ as $H_{\bm k}$ and the valley Chern number $\mathcal C_{\bm \kappa}$ as Chern number $\mathcal C$ from here as it is understood that the other valley is connected by time-reversal symmetry.

In this work, we consider potentials that are experimentally realizable and are compatible with translational symmetry and certain rotational symmetries. Since we are primarily interested in describing the topological transitions of the lowest-energy valence band which are experimentally accessible, we can further constrain the potential by only considering the first and second star of Fourier components. We will see that this is sufficient as they directly couple the nearest high symmetry points and can provide necessary band inversions to obtain non-trivial topologies. Since valence bands have negative energies, we use the term lowest-energy valence band to be associated with the magnitude of energy or closest to the Fermi level at zero.


\section{The $C_4$ symmetric Potential}
\label{sec:C4}

\begin{figure}
\includegraphics[width=\columnwidth]{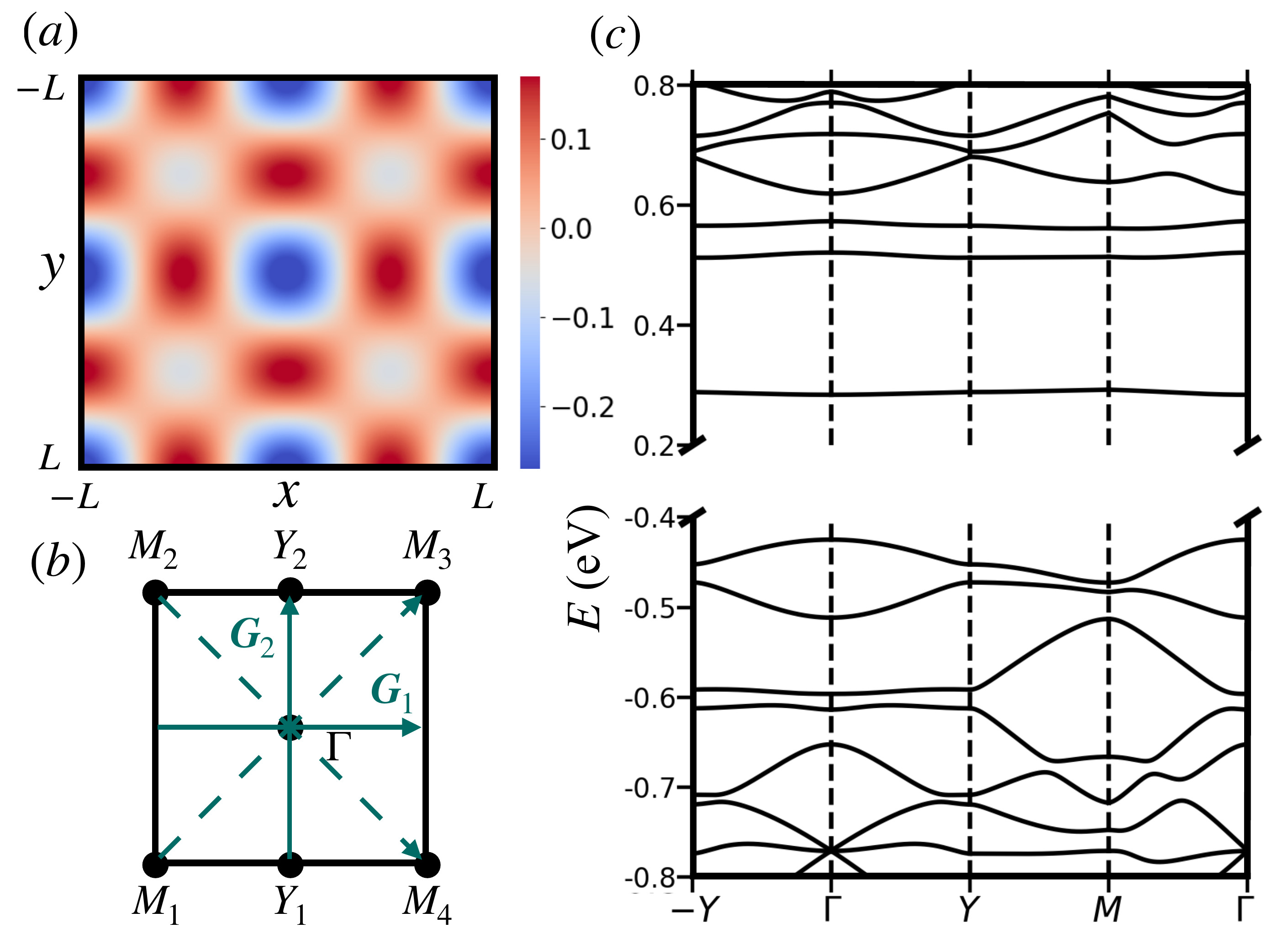}
\caption{ (a) The scalar potential $V(\bm r)$ in real space is plotted for $V_1=0$, $V_0 = 0.14~\text{eV}$, $V_0^\prime = -0.15~\text{eV}$. (b) The first Brillouin zone of the periodic $C_4$ symmetric potential indicating the high symmetry points, with the green solid and dashed arrows marking the first and second star of reciprocal lattice vectors respectively. (c) The valence and conduction bands for the Hamiltonian with $\Delta = 1~\text{eV}$, $t=0.7~\text{eV}$, $a=3.56~\angstrom$ and $L = 10 a$. 
}
\label{fig:C4}
\end{figure}

The most general form of a $C_4$ symmetric scalar potential is $V_0(\bm r) = {1}/{2}\sum_{\bm G} V_0(\bm G) e^{i \bm G \cdot \bm r}$, where $\bm G$ are the set of the star of Fourier vectors connected by $C_4$ symmetry and $V_0(\bm G)$ are real. We will first show that it is necessary to consider both the first and second star of Fourier components to induce $\mathcal C=\pm 1$ for the valence band. We will then go on to show that $\mathcal C= \pm 2$ can only be realized in the valence band with a minimal potential having the scalar part together with the non-scalar part that couples to $\sigma_z$. We consider the $C_4$ symmetric potential given by 
\begin{align}
V(\bm r) = (V_0 I + V_1 \sigma_z) \sum_{i=1}^2 \cos(\bm G_i \cdot \bm r) + V_0^\prime I \sum_{i=3}^4 \cos(\bm G_i \cdot \bm r)\;, \label{eqn:C4_pot}
\end{align}
such that the reciprocal lattice vectors are
$\bm G_1 = 2\pi/L \; [1,0]^T, \; \bm G_2 = 2\pi/L \; [0,1]^T, \; \bm G_{3,4} = \bm G_1 \pm \bm G_2, $ 
and $L$ is the period of the applied external potential. The real coefficients $V_0, V_0^\prime$ correspond to the first (${\bm G_{1,2} }$) and second ($\bm G_{1}\pm \bm G_{2}$) stars of the scalar potential marked by green solid and dashed lines in Fig.~\ref{fig:C4} (b) respectively. The coefficient $V_1$ corresponds to the strength of the non-scalar part of the potential. 

For a $C_4$ symmetric Hamiltonian, the Chern number $\mathcal C_j$ for the $j^{\text{th}}$ band can be calculated by~\cite{PhysRevB.86.115112}
\begin{align}
i^{\mathcal C_j} = \xi_j(\bm \Gamma) \xi_j(\bm M) \zeta_j(\bm Y)\;,
\label{eqn:Chern_C4}
\end{align}
where $\xi_j(\bm \Gamma),\xi_j(\bm M)$ are the eigenvalues of the $C_4$ operator at points $\bm \Gamma,\bm M$ respectively, and $\zeta_j(\bm Y)$ is the eigenvalue of the $C_2$ operator at $\bm Y$ point for the $j^{\text{th}}$ energy band. The total Hamiltonian $H_{\bm k}$ commutes with the rotational operators at respective high symmetry points such that $[C_4,H_{\bm k}]=0$ for $\bm k \in \{ \bm \Gamma, \bm M\}$ and $[C_2, H_{\bm Y}]=0$, and therefore has common non-degenerate eigenvectors. We find the common eigenvectors by diagonalizing $H_{\bm k}$ and use them to find the corresponding eigenvalues of the rotational operators. 
Compared to the direct calculation of the Chern number by numerically integrating the Berry curvature over the Brillouin zone, symmetry analysis provides a great inexpensive and insightful method that can be used to engineer Chern numbers.

We write the Hamiltonian using degenerate perturbation theory by considering sufficient number of Bloch vectors to achieve convergence in describing the low energy physics. For the lowest energy valence band it suffices to take Bloch vectors within the first Brillouin zone at $\bm \Gamma$, $\{\bm Y_1, \bm Y_2\}$ and $\{ \bm M_1, \bm M_2, \bm M_3, \bm M_4\}$ as shown in the inset of Fig~\ref{fig:C4}~(b). We can further simplify the calculation by only considering the valence band states which are of our interest and neglect the conduction bands as $\Delta \gg V_0, V_0^\prime, V_1$. 

We now express the potential in the basis $\ket{u_{\bm k}}$ which are the eigen Bloch vectors of the Dirac Hamiltonian $H^D_{\bm k}$ such that $H^D_{\bm k}\ket{u_{\bm k}}= -\sqrt{\Delta^2/4+(atk)^2} \ket{u_{\bm k}}$ at the respective high symmetry points. 
At the $\bm \Gamma$ point, we only consider the Bloch vector at a single rotationally invariant point and therefore $\xi(\bm \Gamma) = 1$. At the $\bm Y$ point, we express the matrix potential in the basis $\{ \ket{u_{\bm Y_1}},\ket{u_{\bm Y_2}}\}$ to obtain
\begin{align}
V_{\bm Y} = \begin{bmatrix}
				0 & u \\
				u & 0
			\end{bmatrix}\;,
\end{align}
such that $u = \bra{u_{\bm Y_1}}V\ket{u_{\bm Y_2}}= \frac{V_0}{2\sqrt{1+s/2}} - \frac{V_1}{2}$, where $s = 8\pi^2 t^2 a^2/(L\Delta)^2$. We diagonalize it to obtain the eigenenergies
\begin{align}
E_{\pm 1} &=\pm \bigg(\frac{V_0}{2} \frac{1}{\sqrt{1+s/2}} -\frac{V_1}{2}\bigg)\;,
\label{eqn:E_y}
\end{align}
which are labeled by their $C_2$ eigenvalues. Therefore we calculate $\zeta(\bm Y) = \lambda $, where $E^{\bm Y}_\lambda=\max\{ E^{\bm Y}_{1},E^{\bm Y}_{-1}\}$. Similarly at $\bm M$ point we express the matrix potential in the basis $\{ \ket{u_{\bm M_1}}, \ket{u_{\bm M_2}}, \ket{u_{\bm M_3}}, \ket{u_{\bm M_4}} \}$ to obtain
\begin{align}
V_{\bm M} = \begin{bmatrix}
0 & w & v & w^* \\
w^* & 0 & w & v \\
v & w^* & 0 & w\\
w & v & w^* & 0
\end{bmatrix}\;,
\end{align}
where $w = \bra{u_{M_1}}V \ket{u_{M_2}} = \frac{V_0}{2} ( \frac{1+i}{2}) ( 1-\frac{i}{\sqrt{1+s}})+ \frac{V_1}{2}(\frac{1}{2}+\frac{i}{2}) (i-\frac{1}{\sqrt{1+s}})\;,$ and 
$v = \bra{u_{M_1}}V\ket{u_{M_3}} =  \frac{V_0^\prime}{2\sqrt{1+s}}$. We diagonalize it to obtain to obtain the eigenenergies
\begin{align}
E_{\pm 1}&=  +\frac{V_0^\prime}{2\sqrt{1+s}}\pm\frac{1}{2}(V_0-V_1)\bigg ( 1+\frac{1}{\sqrt{1+s}}\bigg)\;,\nonumber \\
E_{\pm i}&= -\frac{V_0^\prime}{2\sqrt{1+s}}\pm\frac{1}{2}(V_0+V_1)\bigg ( 1-\frac{1}{\sqrt{1+s}}\bigg)\;, \label{eqn:E_m}
\end{align}
which are similarly labeled by their $C_4$ eigenvalues. Similarly we get $\xi(\bm M) = \lambda$ such that $E^{\bm M}_{\lambda} = \max \{ E^{\bm M}_{1},E^{\bm M}_{-1},E^{\bm M}_{i},E^{\bm M}_{-i}\}$. 

\subsection{Scalar Potential}

\begin{figure}
\includegraphics[width=\columnwidth]{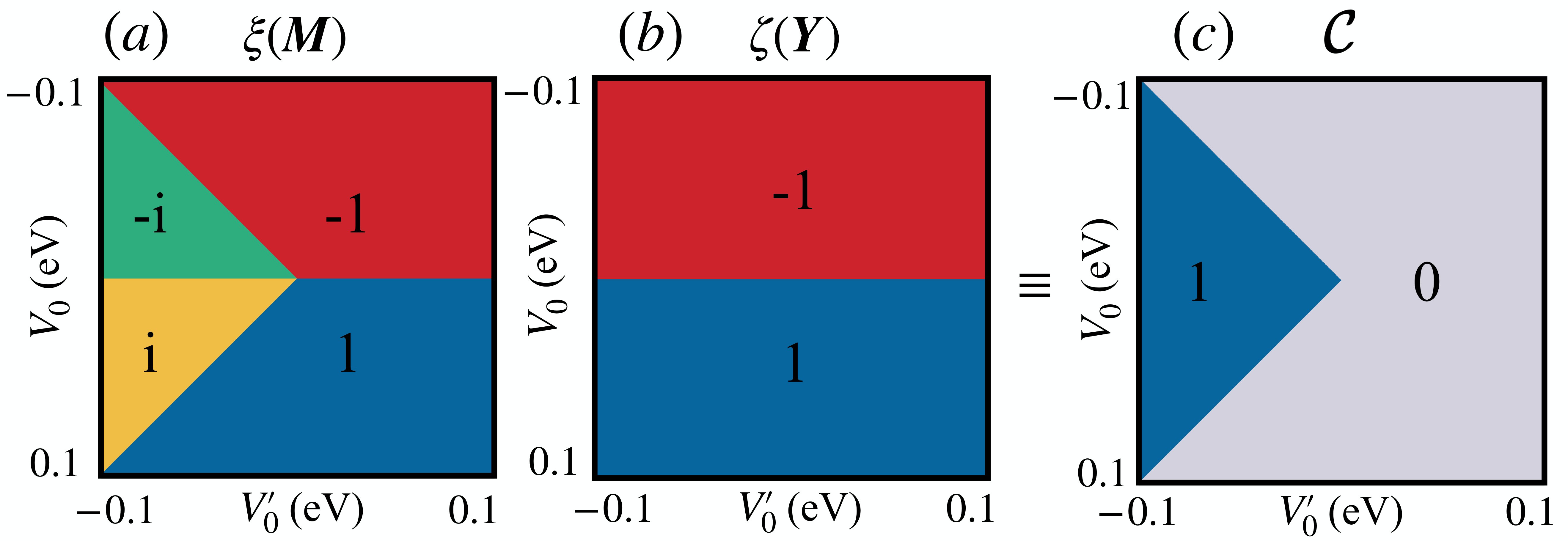}
\caption{(a) The $C_4$ eigenvalues denoted by $\xi(\bm M)$ and (b) the $C_2$ eigenvalues $\zeta(\bm Y)$ of the lowest energy valence band as a function of $V_0,V_0^\prime$, where $V_0 \neq 0$. We know that $\xi(\bm \Gamma)=1$, then Eq.~\ref{eqn:Chern_C4} is used to calculate (c) the Chern number $\mathcal{C}$ for the lowest energy valence band.}
\label{fig:Scalar_C4}
\end{figure}

We first analyze the case when the external potential acts as a scalar by setting $V_1 = 0$. This is typically easier to achieve as the applied field varies slowly over multiple monolayer unit cells and couples to the two orbitals identically. We see that the second star Fourier component of the potential $V_0^\prime$ is necessary to break the degeneracy between $E_{1} (E_{-1})$ and $E_{i} (E_{-i})$ at $\bm M$ point as can be seen from Eq.~\ref{eqn:E_m}. We  now show that it is also essential to induce topological transitions with $\mathcal C = \pm 1$. To see this, let us analyze the condition for band inversion given by Eq.~\ref{eqn:Chern_C4} that implies $\xi(\bm M) \zeta(\bm Y) = \pm i$. Using Eq.~\ref{eqn:E_y} and Eq.~\ref{eqn:E_m}, it can be shown that $V_0^\prime < - V_0$ or $|V_0^\prime|>|V_0|$ such that $\text{sign}(V_0^\prime)=-1$. 
Therefore, not only is the second star of the Fourier component essential to realize a non-trivial topological transition, it must be bigger than the magnitude of the first star of the Fourier component of the potential. Typically, this is not possible in naturally occurring moir\'e potentials which are well described by only the first order component where $V_0^\prime=0$. Therefore our result showcases the unique freedom provided by the externally applied potential that is not found in potentials created by the intrinsic or natural moir\'e pattern.

As an example, we take values typical of TMD monolayers by choosing $\Delta = 1~\text{eV}$, $t=0.7~\text{eV}$, and $a=3.56 \angstrom$. We take the external scalar potential $V(\bm r)$ parameterized by $V_0 = 0.14~\text{eV},$ $V_0^\prime = -0.15~\text{eV}$, and $L = 10a$ shown in Fig.~\ref{fig:C4} (a). To obtain the band structure, we represent the Hamiltonian using a sufficient number of Bloch vectors for convergence and solve the central equation to obtain the valence and conduction bands shown in Fig.~\ref{fig:C4} (c). The non-trivial coupling $V(\bm r)$ lifts the degeneracy of the Dirac bands at high symmetry points. 
In Fig.~\ref{fig:Scalar_C4} we calculate the Chern number for the lowest-energy valence band using symmetry analysis. For each $V_0, V_0^\prime$, we calculate the band inversions at $\bm Y$ and $\bm M$ points. We use Eq.~\ref{eqn:E_m} and Eq.~\ref{eqn:E_y} to obtain the eigenvalues $\xi(\bm M)$ and $\zeta(\bm Y)$ shown in Fig.~\ref{fig:Scalar_C4} (a) and (b) respectively. The Chern number $\mathcal C$ shown in part (c) is then obtained using Eq.~\ref{eqn:Chern_C4}. We see that the non-trivial Chern number arises when $|V_0^\prime|>|V_0|$ which is consistent with our discussion above and have also verified the same numerically by diagonalizing the Hamiltonian to obtain its eigenvectors and using the standard formula of integrating the Berry curvature over the first Brillouin zone.

We can also immediately see that the Chern number $\mathcal C = 2$ is not possible to obtain with the scalar potential of the above form. We can obtain the condition for $\mathcal C=2$ given by $\xi(\bm M) \zeta(\bm Y)=-1$ from Eq.~\ref{eqn:Chern_C4}. Considering the case when $\xi(\bm M) = 1$ and $\zeta(\bm Y) = -1$, we obtain $V_0<0$ from Eq.~\ref{eqn:E_y} and $V_0>0$ from Eq.~\ref{eqn:E_m} leading to a contradiction as both equations cannot be satisfied simultaneously. We can similarly show that the other case with $\xi(\bm M) = 1, \zeta(\bm Y) = -1$ also leads to the same contradiction.

\subsection{Scalar and Non-Scalar Potential}

\begin{figure}[t]
\includegraphics[width=\columnwidth]{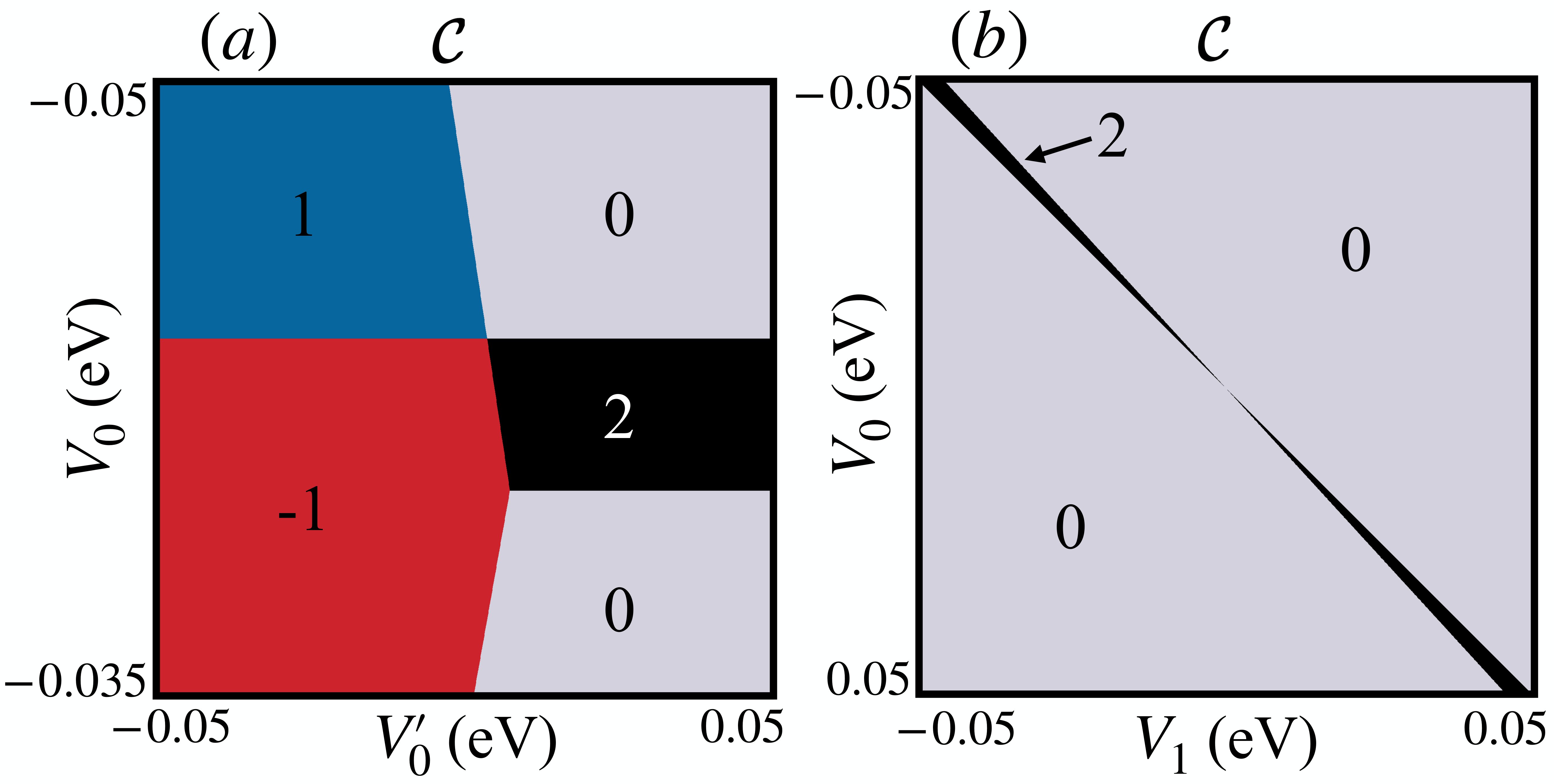}
\caption{The Chern number $\mathcal C$ of the valence band taking (a) $V_1=-0.04~\text{eV}$ and (b) $V_0^\prime=0.05~\text{eV}$.}
\label{fig:Chern2}
\end{figure}

Figure~\ref{fig:Scalar_C4} shows that both $\xi(\bm M), \zeta(\bm Y)$ together undergo band inversion when $V_0$ changes sign and in turn can never realize the condition $\xi(\bm M) \zeta(\bm Y)=-1$ for $\mathcal C=2$. Motivated by this insight and realizing that this band inversion symmetry must be broken, we introduce the non-scalar term $V_1$ that affects $\bm M, \bm Y$ points differently as confirmed from Eq.~\ref{eqn:E_y},~\ref{eqn:E_m}. We show in Fig.~\ref{fig:Chern2} that the non-scalar term along with the scalar potential can obtain $\mathcal C=2$. 

In fact, we remark that the $C_4$ symmetric potential $V(\bm r)$ given in Eq.~\ref{eqn:C4_pot} is the minimal form necessary to obtain $\mathcal C=2$. To see this, we analyze the case when $\xi(\bm M)=1$ and $\zeta(\bm Y)=-1$, then Eq.~\ref{eqn:E_y},~\ref{eqn:E_m} together imply
\begin{align}
V_1 > \frac{V_0}{\sqrt{1+s}}, \; V_0 > V_1,\; V_0 >\frac{V_1-V_0^\prime}{\sqrt{1+s}},\;  \frac{V_0^\prime + V_0}{\sqrt{1+s}}>V_1\;.
\end{align}
If we take $V_0^\prime=0$, we obtain the constraints $V_0/\sqrt{1+s}>V_1$ and $V_0/\sqrt{1+s}<V_1$ which is a contradiction. Despite the non-scalar term, we see that the second Fourier component $V_0^\prime$ which does not occur in natural moir\'e potentials is necessary to induce $\mathcal C=2$. Similarly, if we chose $V_0=0$, we obtain the constraints $V_1<0$ and $V_1>0$ thus obtaining another unsatisfiable condition. It is then evident that the equations can only be simultaneously satisfied when all three terms with $V_0, V_0^\prime,$ and $V_1$ are present. Therefore, the potential given in Eq.~\ref{eqn:C4_pot} has the minimal form to obtain $\mathcal C=\pm 2$.

\section{The $C_3$ symmetric Potential}
\label{sec:C3}
In this section, we analyze the $C_3$ symmetric potential and show that $\mathcal C=2$ can be realized similarly by having a scalar together with a non-scalar periodic potential.
It is worthwhile to note that the symmetry analysis method applied to a system with symmetry $C_n$ can only predict Chern number modulo $n$. 
Therefore for the $C_4$ system, the symmetry method could not distinguish between $\mathcal C = \pm 2$. Similarly, for a $C_3$ system, it cannot distinguish between $\mathcal C=2,-1$. We nevertheless show in this section that it is in fact possible to infer when the $\mathcal C=2$ phase transition happens indirectly.  We achieve this by using the symmetry method that takes into account the band inversions at high $C_3$ symmetric points $\bm K,-\bm K$, together with analyzing other points that are connected by the reciprocal lattice vectors of the applied potential that can also cause band inversions.

We write the potential with a scalar and a non-scalar coupling upto the first star of Fourier vectors that is allowed by $C_3$ symmetry as
\begin{align}
V(\bm r) = V_0 \sum_{i=1}^3\cos(\bm G_i\cdot \bm r + \phi_0) + V_1 \sigma_z \sum_{i=1}^3\cos(\bm G_i\cdot \bm r + \phi_1)\;,
\label{eqn:C3_pot}
\end{align}
where $\bm G_{1,2} = \frac{2\pi}{3L} [1, \pm \sqrt{3}]^T$, $\bm G_3 = - \bm G_1 - \bm G_2$ are the reciprocal lattice vectors of the Honeycomb lattice, $V_0$, and  $V_1$ are the strengths of the scalar and non-scalar parts of the periodic potentials respectively. Reference~\cite{su2021massive} considered the scalar part of the potential that naturally occurs in MoTe$_2$/WSe$_2$ TMD heterobilayer. 

\begin{figure}
\centering
\includegraphics[width=\columnwidth]{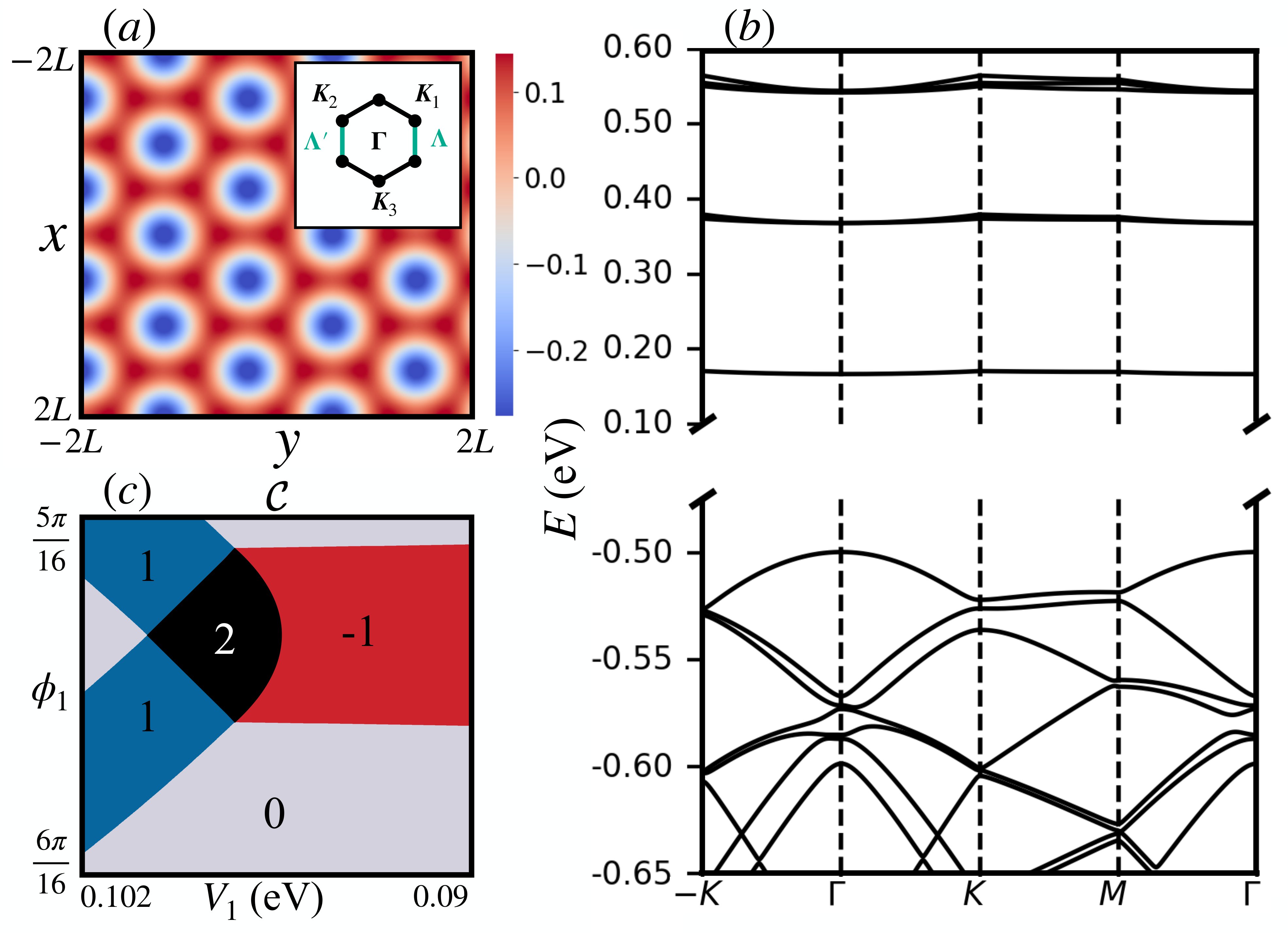}
\caption{(a) The scalar part of the potential $V(\bm r)$ in Eq.~\ref{eqn:C3_pot} with $V_0 = 0.1~\text{eV}$ and $\phi_0=\pi/3$. The inset shows the first Brillouin zone created by the $C_3$ symmetric potential indicating high symmetry points and lines. The non-scalar part (not shown) is taken as $V_1 = 0.099~\text{eV}$ and $\phi_1 = 5.5\pi/16$. (b) The valence and conduction bands for the Hamiltonian with $\Delta = 1~\text{eV}$, $t=0.7~\text{eV}$, $a=3.56~ \angstrom$ and $L = 10 a$.  (c) Chern number for the lowest-energy valence band by numerical integration of the Berry curvature over the Brillouin zone.}
\label{fig:C3_bands}
\end{figure}

We consider an example by taking values typical of TMD monolayers by choosing $\Delta = 1~\text{eV}$, $t=0.7~\text{eV}$, and $a=3.56 \angstrom$. We take $C_3$ symmetric potential $V(\bm r)$ parameterized by $V_0 = 0.1~\text{eV}$, $\phi_0 = \pi/3$, $V_1=0.099~\text{eV}$, $\phi_1 = 5.5\pi/16$ and $L=10a$ where the scalar part is shown in Fig.~\ref{fig:C3_bands} (a) and the first Brillouin zone shown in the inset. Figure~\ref{fig:C3_bands} (b) shows the band structure for the total Hamiltonian with the valence and conduction bands. In Fig.~\ref{fig:C3_bands} (c), we vary $V_1,\phi_1$ and obtain the Chern number $\mathcal C$ for the lowest energy valence band numerically by integrating the Berry curvature over the Brillouin Zone. We, therefore, present the potential parameters that can be tuned to achieve the $\mathcal C=2$ phase. However, this purely numerical exercise is not only expensive but offers no insight into finding the potential parameters to obtain $\mathcal C=2$, which is very difficult even when the potential is parameterized with only five parameters as in this case. In fact, we found the relevant parameters using the symmetry method described below.

\begin{figure}
\centering
\includegraphics[width=\columnwidth]{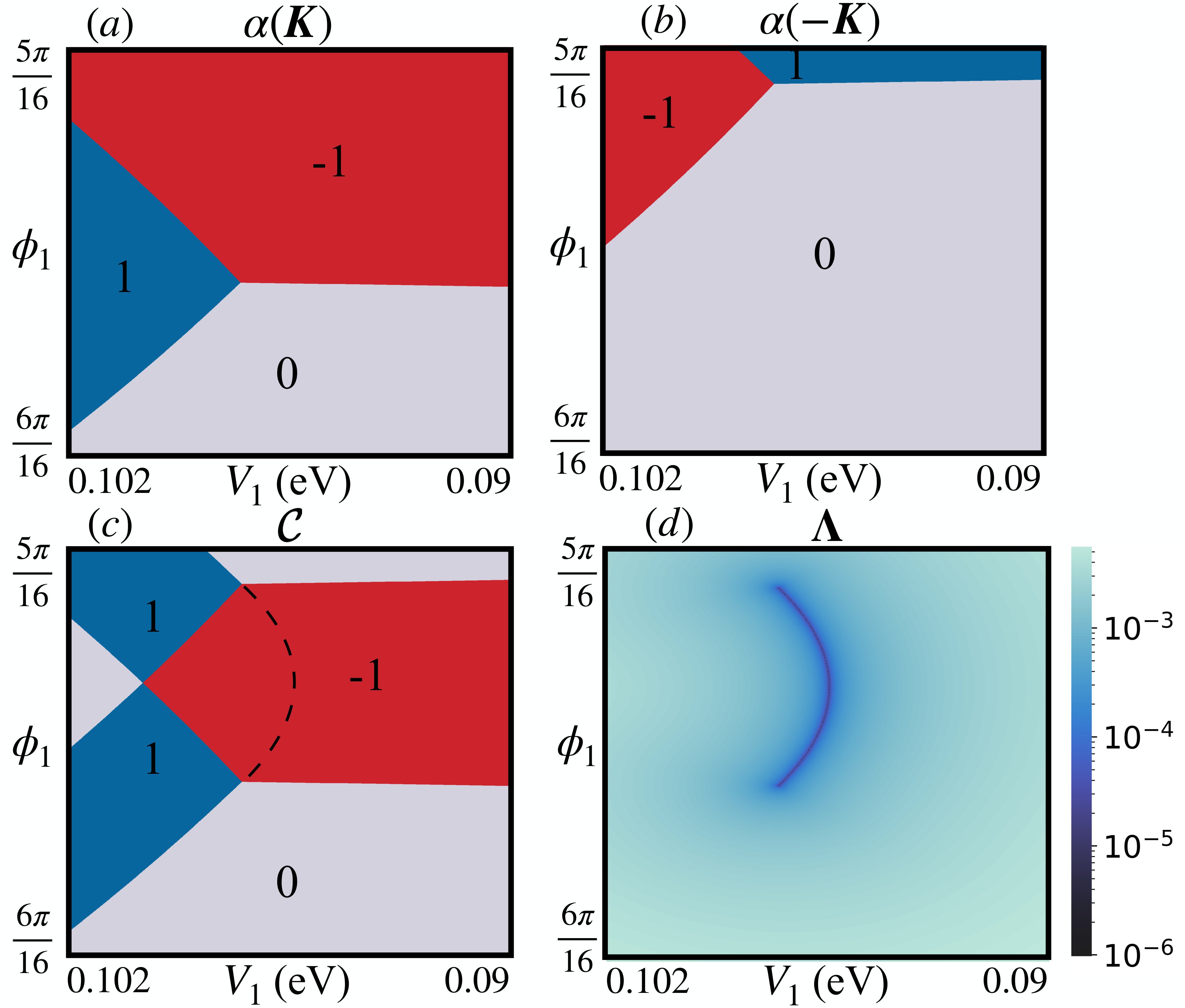}
\caption{The Hamiltonian is parameterized with $\Delta = 1~\text{eV}$, $t=0.7~\text{eV}$, and $a=3.56~ \angstrom$. The potential parameters $V_0=0.1~\text{eV}, \phi_0=\pi/3$, and $L=10a$ are fixed. The phase (a) $\alpha(\bm K)$, (b) $\alpha(-\bm K)$, and Chern number (c) $\mathcal C$ of the lowest energy valence band as a function of tunable potential parameters $V_1,\phi_1$. (d) The minimum band gap along the line $\bm \Lambda$ given by $\min_\beta \delta(\beta)$ as a function of $V_1,\phi_1$.}
\label{fig:C3_symm}
\end{figure}

For a $C_3$ symmetric Hamiltonian, the Chern number $\mathcal C$ can be calculated up to modulus $3$ by using symmetry analysis~\cite{PhysRevB.86.115112, su2021massive}
\begin{align}
e^{i\frac{2\pi}{3}\mathcal C_j} = \eta_j(\bm \Gamma)\eta_j(\bm K)\eta_j(-\bm K) \;,
\label{eqn:Chern_C3}
\end{align}
where $\eta_j(\bm k)$ denotes the $C_3$ eigenvalue of the corresponding band $j$, at the high symmetry point $\bm K$. For the lowest energy valence band, it suffices to consider Bloch vectors within the first Brillouin zone at $\bm \Gamma$, and $\pm \boldsymbol K = \{ \pm \bm K_1, \pm \bm K_2, \pm \bm K_3\}$ as shown in the inset of Fig.~\ref{fig:C3_bands} (a). We similarly simplify our calculation by only considering the valence bands by working in the regime where $\Delta \gg V_0, V_0^\prime, V_1$. 

We again express the potential in the basis $\ket{u_{\bm k}}$ which are the eigen Bloch vectors of the Dirac Hamiltonian $H^D_{\bm k}$. At the $\bm \Gamma$ point, we only consider the Bloch vector at a single rotationally invariant point and therefore $\eta(\bm \Gamma) = 1$.
At $\pm \bm K$ set of points, we write the matrix $V_{\pm\bm K}$ in the basis of the Bloch vectors $\{ \ket{u_{\pm\bm K_1}},\ket{u_{\pm\bm K_2}},\ket{u_{\pm\bm K_3}}\}$ as
\begin{align}
V_{\pm \bm K} = \begin{bmatrix}
				0 & v(\pm \phi) & v^*(\pm \phi)\\
				v^*(\pm \phi) & 0 & v(\pm \phi)\\
				v(\pm \phi) & v^*(\pm \phi) & 0 
				\end{bmatrix}\;,
\end{align}
where the overlap $v(\pm \phi)  = \bra{u_{\bm K_1}}V\ket{u_{\bm K_2}}$ is
\begin{align}
v(\pm \phi) 
&=\frac{V_0}{2} e^{i(\pm\phi_0+\pi/3)} \bigg( \frac{1}{2}- \frac{i\sqrt{3}}{2\sqrt{1+s^\prime}}\bigg) \nonumber \\ &+ \frac{V_1}{2} e^{i(\pm\phi_1+\pi/3)}\bigg( i\frac{\sqrt{3}}{2} - \frac{1}{2\sqrt{1+s^\prime}}\bigg) \;,
\end{align}
where $s^\prime = 64\pi^2 t^2 a^2/(27 L^2 \Delta^2)$.
The eigenvalues of the above matrix are
\begin{align}
E_0 &= 2\Re(v)\; & E_{\pm 1}&= -\Re(v) \pm \sqrt{3}\Im(v) \;,
\end{align}
with corresponding eigenvectors $\ket{E_\alpha}$, such that $C_3 \ket{E_\alpha} = e^{-i\alpha2\pi/3} \ket{E_\alpha}$, where $\alpha \in \{ 0,\pm 1\}$. We can now find $\eta(\boldsymbol{K}) = e^{-i\alpha(\bm K) 2\pi/3}$ by taking $v=v(+\phi)$ and finding $\alpha(\bm K): E_\alpha = \max\{ E_0, E_1, E_{-1}\}$ and take $v=(-\phi)$ to similarly obtain $\eta(-\boldsymbol K)$. We can obtain the Chern number $\mathcal C$ modulo three of the lowest energy valence band by using Eq.~\ref{eqn:Chern_C3}. Figure~\ref{fig:C3_symm} (a),(b) shows $\alpha (\pm\bm K)$ as a function of tunable potential parameters $V_1, \phi_1$ and (c) shows the Chern number $\mathcal C$ for the lowest energy valence band as calculated from Eq.~\ref{eqn:Chern_C3}. 

Since the $\mathcal C=2$ phase is rare and non-trivial, finding the potential parameters to engineer it requires being able to reliably distinguish between $\mathcal C= 2, -1$. We realize that the external $C_3$ potential is also mirror-symmetric from Fig.~\ref{fig:C3_bands} (a), and thus the high symmetry lines $\bm \Lambda, \bm \Lambda^\prime$ shown in green in the inset are also connected by a reciprocal lattice vector of the applied potential. Therefore, varying the potential parameters can in principle also cause a band inversion anywhere along $\bm \Lambda$ which is an indicator of a phase change between $\mathcal C=2,-1$. We can calculate the band inversions along $\bm \Lambda$ by parameterizing the lines $\bm \Lambda,\bm \Lambda^\prime:$ $\bm \Lambda(\beta) = -\bm K_2 + \beta (\bm K_1 + \bm K_2)$ and $\bm \Lambda^\prime(\beta) = -\bm K_1 + \beta (\bm K_2 + \bm K_1)$, where $0<\beta<1$ and writing the potential $V(\bm r)$ in the basis of Bloch vectors $\{ \ket{u_{\bm \Lambda(\beta)}}, \ket{u_{\bm \Lambda^\prime (\beta)}}\}$ as 
\begin{align}
V_{\bm \Lambda(\beta)} = \begin{bmatrix}
                    0 & \delta(\beta) \\ 
                    \delta^*(\beta) & 0 \;
                    \end{bmatrix}
\end{align}
where $\delta (\beta) = \bra{u_{\bm \Lambda(\beta)}}V\ket{u_{\bm \Lambda^\prime(\beta)}}$ and can be calculated analytically with its form given in the footnote~\footnote{ The matrix element $\delta (\beta) = \bra{u_{\bm \Lambda(\beta)}}V\ket{u_{\bm \Lambda^\prime(\beta)}}$ is given by  
$ {\scriptstyle
\delta(\beta) = \frac{V_0}{2}e^{i\phi_0} \frac{s^\prime/4(2\beta-1)(2\beta-1+i\sqrt{3})+ 1 + \sqrt{s^\prime(1+(\beta-1)\beta) + 1}}{|s^\prime(1+(\beta-1)\beta) + 1 + \sqrt{s^\prime(1+(\beta-1)\beta) + 1}|}} \\{\scriptstyle 
+\frac{V_1}{2}e^{i\phi_1} \frac{is^\prime/4 (3i-\sqrt{3}+2\sqrt{3}\beta) - 1 - \sqrt{s^\prime(1+(\beta-1)\beta) + 1}}{|s^\prime(1+(\beta-1)\beta) + 1 + \sqrt{s^\prime(1+(\beta-1)\beta) + 1}|}\;}
$}. The eigenvalues of the matrix are $ \pm |\delta(\beta)|$ and there is a band closing at point $\bm \Lambda(\beta)$ when $\delta(\beta) = 0$. In our example, we plot the minimum band gap along $\bm \Lambda$ given by $\min_{\beta} \delta(\beta)$ as a function of $V_1, \phi_1$ in Fig.~\ref{fig:C3_symm} (d) and notice that the black line represents band closings somewhere along $\bm \Lambda$. Therefore, we can determine an additional crossing at $\bm \Lambda$ which was not captured by Eq.~\ref{eqn:Chern_C3} missing the Chern transition between $\mathcal C=2,-1$. We represent this crossing with a dashed line in Fig.~\ref{fig:C3_symm} (c) which dovetails the numerical result in Fig.~\ref{fig:C3_bands} (c).

\section{Experimental Realization}
\label{sec:Exp}
We saw that for both $C_4, C_3$ symmetric potentials, the scalar part is enough to induce $\mathcal C=1$ for one valley. This opens the door to realize QSHE as TMDs have strong spin-orbit coupling with the other valley having opposite spin and Chern number. 
As mentioned in the introduction, experimental realizations of this can be achieved with devices created by placing a monolayer TMD on either an LAO/STO substrate, a twisted h-BN substrate or by considering dielectric superlattices. We focus our discussion on the LAO/STO example in this section.
A superlattice potential is induced by using an electron beam to write conductive nanostructures at the insulating LAO/STO interface \cite{yang2020nanoscale} for substrates having 3 layers of LAO.
This experimental system can be used to create the external scalar superlattice potential with the first and second Fourier components as shown in Fig.~\ref{fig:C4} (a).
The terms $V_0$ and $V_0'$ in the scalar potential can, in principle, be achieved by
writing the two regions in Fig.~\ref{fig:C4} (a) using different doses of the electron beam. 
This external potential created by the LAO/STO will couple identically to both {sublattices in the TMD monolayer.} Poisson's equation may be solved to obtain the exact charge density required at the LAO/STO interface, which can be tuned using a back gate. 

We saw that for both $C_4$ and $C_3$ symmetric potentials, the scalar together with a non-scalar part is necessary to induce non-trivial topological transitions such as $\mathcal C=2$. It may be hard to introduce a non-scalar $C_4$ symmetric coupling, but in principle may be possible by strong corrugation or sandwiching the TMD between two external potentials with LAO/STO setup on top and bottom. For the $C_3$ symmetric potential, the non-scalar periodic coupling might be easier to realize as it is naturally present in bilayer graphene or graphene-hBN bilayer. 

Making bands flat with a non-trivial Chern number such as $\mathcal C=2$ will lead to strongly correlated phenomena and FQHE physics potentially producing non-Abelian anyons not possible with an external magnetic field. Bands can be made flatter by increasing the coupling strength of the potential and the periodic length. Taking the charge density at the interface of LAO/STO to be $\sigma = 2\times 10^{13} ~\frac{e}{\text{cm}^2}$ and approximating the 2-DEG as an infinite charged sheet $V_0 = \frac{\sigma d}{2\epsilon_0}$, where $d=1.2~\text{nm}$, the length of $3$ LAO layers and $\epsilon_0$ is the electric vacuum permittivity. We find the rough estimate $V_0\sim 2~\text{eV}$ which should be strong enough to produce flat bands. This is possibly the upper limit as the 2DEG may not be very large as compared to the TMD.  

In summary, we considered Dirac fermions under externally applied potentials that can be created and controlled using contemporary experimental techniques. We analyzed a class of $C_4$ amd $C_3$ symmetric potentials and provided their minimal forms to realize non-trivial topological phases of matter, discovering non-trivial transitions that may be difficult to realize with natural moir\'e potentials. We showed that $\mathcal C=2$ can be realized for both classes of potentials by considering a scalar together with a non-scalar coupling based on symmetry analysis and provided numerical verification. Our work opens up the possibility to realize QSHE, QHE, and even non-Abelian Anyons in FQHE physics using current experimental hardware.

\acknowledgments{This work is supported by the Department of Energy BES QIS program on `Van der Waals Reprogrammable Quantum Simulator' under award number DE-SC0022277. NS also acknowledges support from NASA Academic Mission Services contract NNA16BD14C. }

\bibliography{Engineer_Topology}

\end{document}